\PassOptionsToPackage{numbers,square,sort&compress}{natbib} 
\documentclass[review]{elsarticle}
\usepackage{amsmath,amssymb,amsbsy,lineno,hyperref,subcaption}
\usepackage{tabularx, tikz, xcolor, mathtools, tabu, pifont, gensymb, wrapfig}
\usepackage{ragged2e, amsmath, geometry, blindtext, wasysym, adjustbox, array}
\usepackage[T1]{fontenc} 
\justifying
\newcommand{\baln}{B$_{x}$Al$_{1-x}$N }
\newcommand{\algan}{Al$_{x}$Ga$_{1-x}$N }
\newcommand{\balnend}{B$_{x}$Al$_{1-x}$N}
\begin{document}

\begin{frontmatter}

\title{Electronic Properties of Ultra-Wide Bandgap B$_x$Al$_{1-x}$N Computed from First-Principles Simulations}

\author{Cody L. Milne*}
\author{Tathagata Biswas}
\author{Arunima K. Singh}
\ead{clmilne@asu.edu, tbiswas3@asu.edu, arunimasingh@asu.edu}
\address{Department of Physics\\
Arizona State University\\
Tempe, AZ 85282\\}

\begin{abstract}
Ultra-wide bandgap (UWBG) materials such as AlN and BN hold great promise for future power electronics due to their exceptional properties. They exhibit large bandgaps, high breakdown fields, high thermal conductivity, and high mechanical strengths. AlN and BN have been extensively researched, however, their alloys, \balnend, are much less studied despite their ability to offer tunable properties by adjusting $x$. In this article, we predict the electronic properties of 17 recently predicted ground states of \baln in the $x=0-1$ range using first-principles density functional theory and many-body perturbation theory within $GW$ approximation. All the \baln structures are found to be UWBG materials and have bandgaps that vary linearly from that of wurtzite-phase (\emph{w}) AlN (6.19 eV) to that of \emph{w}-BN (7.47 eV). The bandstructures of \baln show that a direct-to-indirect bandgap crossover occurs near $x = 0.25$. Furthermore, we find that \baln alloys have much larger dielectric constants than the constituent bulk materials (AlN=$9.3~\varepsilon_0$ or BN=$7.3~\varepsilon_0$), with values reaching as high as $12.1~\varepsilon_0$. These alloys are found to exhibit large dielectric breakdown fields in the range 9--35 MV/cm with a linear dependence on $x$. This work provides the much needed advancement in the understanding of the properties of \baln to aid their application in next-generation devices.
\end{abstract}


\end{frontmatter}

\section{Introduction}

Ultra-wide bandgap (UWBG) materials are traditionally defined as materials that have a bandgap larger than that of GaN (3.4 eV) \cite{uwbg-semiconductors}. Recently, these materials have received burgeoning interest due to their exciting applications in optoelectronics, radio frequency devices, and high-voltage/power electronics. Many performance parameters depend heavily on the bandgap, $E_g$, for example, the drift layer thickness and the specific on-resistance in power electronic devices. In diodes, higher bandgaps lead to a reduction of impact ionization rates and tunneling effects. UWBG materials also display large breakdown fields \cite{Tsao2018, Wong2021, Slobodyan2022}, on the order of MV/cm, that lead to a reduction in leakage currents and are expected to enable significant miniaturization of devices. 

AlN and BN in their wurtzite-phase (\emph{w}-phase) are UWBG materials with bandgaps of 6.2 eV and 5.44--7.70 eV, respectively \cite{Kudrawiec2020}. Therefore \baln alloys are expected to display much higher bandgaps than the widely used \algan alloys \cite{Shen2017,ZhangQ2022,Kudrawiec2020}. In the recent past, \baln has been grown in the thin-film form with boron fractions up to $x=0.30$ and film thicknesses up to 300 nm \cite{Li2015, Li2017, Sun2017, Sun2018, Tran2020, Vuong2020, Hayden2021, Sarker2020, ZhangQ2022, Zhu2021, Wolff2021, Calderon2023}. Additionally, little is known experimentally about the doping characteristics in \baln\cite{Amano2020} since it was only recently synthesized\cite{Li2015, Li2017, Sun2017, Sun2018, Tran2020, Vuong2020, Hayden2021, Sarker2020, ZhangQ2022, Zhu2021, Wolff2021, Calderon2023}, but Be and Mg have been predicted to be promising $p$-type donors\cite{Weston2017} for w-BN. In spite of the recent interest in \emph{w}-\balnend, only a few studies have investigated their structure, electronic, and dielectric properties \cite{Zhang2017, Dong2021, Ilyasov2005, ZhangQ2022, Lu2021, Ota2022}.

In our recent work, we have shown that \cite{Milne2023} \baln can have thermodynamically stable structures in the entire $x=0-1$ range. In that study, we predicted the structure and formation energies of ground-states of \baln alloys in the entire $x=0-1$ range using the density-functional theory (DFT)-based cluster expansion method. This approach allowed the prediction of \baln atomic structures by considering a wider range of structural configurations and space group symmetries \cite{Ghosh2008,Wu2016} in comparison to the previous studies where cation substitution in the $w$-AlN lattice has been the prevalent approach \cite{Shen2017,Zhang2017,ZhangQ2022}. We obtained 17 ground-state structures of \baln that displayed high formation energies, had predominantly $sp^3$ bonded but distorted tetrahedra, and wurtzite-like lattices. 

In this study, we present the \emph{ab initio} computed bandgaps, band structures, effective masses, dielectric constants, and electric breakdown fields of the 17 \baln ground state structures. We compute the bandgaps and band structures using excited state $GW$ simulations, which greatly increases the accuracy of excited state properties by considering the propagation of electrons and holes through the Green's function, $G$, as well as the screened Coulomb interaction, $W$. In this method, DFT eigenvalues and orbitals are used for a first guess of $G$ (i.e. $G_0$) and $W$ (i.e. $W_0$) known as the single-shot approach of $G_0 W_0$ method. Iteratively solving for $G$ and not $W$ is known as the partially self-consistent approach or $GW_0$ method \cite{Stan2009}. Note that DFT simulations that have been employed to study \baln in the past \cite{Lawniczak-Jablonska2000, Ahmed2007, Zhang2017, Dong2021, Ilyasov2005, ZhangQ2022, Lu2021, Ota2022} severely underestimate the bandgap of semiconductors and insulators \cite{Perdew1985}. \baln has also been studied using hybrid functionals that provide better agreement with experimentally measured bandgaps \cite{Shen2017, Moses2011}. However, hybrid functionals rely on empirical parameters (the fraction of exact exchange) and thus parameter-free many-body perturbation theory methods such as $GW$ simulations are best suited to predict the electronic structure of large gap materials like \baln \cite{Punya2012, Biswas2021, Biswas2023}. Our $GW$ simulations show that \baln structures have bandgaps that linearly vary from that of $w$-AlN (6.19 eV) to $w$-BN (7.47 eV). Further, unfolded band structures show a direct-to-indirect transition at $x=0.25$. We find that all the \baln materials display effective masses that are comparable to that of AlN and BN, however, the structure that has $sp^2$ bonded cations displays significantly higher effective masses. We compute static dielectric constants of the alloys using DFT simulations and find that they are highest at intermediate boron fraction, with values reaching as high as $12.1\varepsilon_0$. We observe large bowing parameters of $b \approx -7.2 \varepsilon_0$ and $b \approx -9.8 \varepsilon_0$ for $\varepsilon_{0}^{\perp}$ and $\varepsilon_{0}^{\parallel}$ respectively. We compute the electric breakdown fields of \baln using a machine-learned model \cite{Kim2016} and find that the breakdown fields of the alloys vary linearly from 9 MV/cm to 35 MV/cm. Thus, we show that \baln exhibits tunable and large bandgaps, high dielectric constants, and high breakdown fields. 

\section{Computational methods}
All \emph{ab initio} simulations reported in this study were performed using the Vienna Ab Initio Simulation Package (VASP) \cite{Kresse1, Kresse2, Kresse3, Kresse4} package. Quasiparticle energies of all the \balnend, AlN, and BN were computed using many-body perturbation theory simulations within the $GW$ approximation. To obtain the quasiparticle bandstructures we used wannier interpolation via maximally localized wannier functions as implemented in the Wannier90 \cite{wannier90,mostofi2008wannier90} package. To obtain unfolded band structures, the BandUP code \cite{bandup1,bandup2} was used. We unfolded the electronic bands for each supercell onto the Brillouin Zone (BZ) of its corresponding primitive cell. The unfolded bandstructures were used to study the element and orbital contributions of different bands at high-symmetry $k$-points and to determine the $x$-value at which the direct-indirect transition occurs in \balnend. The effective masses were calculated by fitting parabolas near $\Gamma$, K, and other valence band maxima (VBM) and conduction band minima (CBM) locations in the BZ of unfolded bandstructures. 

A high-throughput workflow package, \emph{py}GWBSE \cite{Biswas2021,Biswas2023} was used to perform the $GW$ calculations. \emph{py}GWBSE enables automated $GW$ calculations, including convergence of the parameters that are specific to the $GW$ calculations such as the plane-wave cutoff for screened Coulomb potential, the number of empty bands included in the self-energy calculation, and the number of imaginary frequency grid points used in the $GW$ calculation. The $GW$ simulations were performed using a plane-wave energy cutoff of 500 eV, a screened Coulomb energy cutoff of 200 eV, 70 imaginary frequency grid points, and a $\Gamma$-centered \emph{k}-grid of 100 \emph{k}-points per \AA$^{-3}$ of the BZ. These screened Coulomb energy cutoff and frequency grid points resulted in bandgaps that converged within 0.1 eV. The number of bands for each structure was converged by doubling from an initial choice based on the number of atoms in the cell until bandgap convergence within 0.1 eV was reached. For partially self-consistent $GW_0$ calculations, we performed iteration of the Green's function $G$ until the bandgap converged to within 0.1 eV. For all of the \baln structures, \emph{w}-AlN, and \emph{w}-BN, convergence was reached on the fourth iteration. 

The static dielectric constants were calculated using density-functional perturbation theory (DFPT) \cite{Baroni1987, gajdovs2006linear, Gonze1997} by incorporating both the electronic as well as ionic contributions. An energy cutoff of 520 eV and 100 \emph{k}-points per \AA$^{-3}$ of the BZ was used for the DFPT calculations. The Projector-Augmented-Wave (PAW) formalism  \cite{Perdew20} and the PBE \cite{Blochl1994, Perdew19, Perdew20} exchange-correlation functional was employed.

The intrinsic breakdown field was computed using a machine-learned (ML) model by Kim et al. \cite{Kim2016} of the form, 

\begin{equation}
    E_{b} = 0.2444 e^{0.315 \sqrt{E_{g}\omega_{\text{cutoff}}}},
    \label{eq:breakdown}
\end{equation}
where $E_{b}$ is the electric breakdown field in units of MV/cm, $E_g$ is the bandgap energy in units of eV, and $\omega_{\text{cutoff}}$ or the phonon cutoff frequency is the maximum phonon frequency at $\Gamma$ in units of THz. This ML model of Kim et al. is based on a least absolute shrinkage and selection operator (LASSO) based least-squares fit \cite{Tibshirani1996} and was trained on 82 insulating and semiconducting materials. These materials included a wide range of experimental bandgaps (0.2 -- 13.6 eV) and \emph{ab initio} computed intrinsic breakdown fields (0.106 -- 48.3 MV/cm). The phonon cutoff frequency and bandgap were found to be the most predictive parameters amongst all the considered features -- including experimental bandgaps, and \emph{ab initio} predicted phonon cut-off frequency, nearest neighbor distance, mean phonon frequency, dielectric constant, density, and bulk modulus. This model from Kim et al. resulted in a $R^{2}$ value of 0.81 for the training set and 0.72 for the test set. In this work, we have used $E_g$ values as obtained from our $GW_{0}$ calculations and the $\omega_{\text{cutoff}}$ were obtained from our previously computed DFT phonon spectra \cite{Milne2023}.

\section{Results and Discussion}
\subsection{Crystal Structure and Tetrahedrality}
In our previous work, the structure of \baln was predicted using the cluster expansions formalism \cite{avdw:atat,avdw:atat2,avdw:emc2,avdw:maps, avdw:mcsqs} for the entire $x=0-1$ range. We found seventeen ground state structures of the \baln with formation energies between 0.11 and 0.25 eV atom$^{-1}$. The phonon spectra of all the \baln showed non-imaginary phonon frequencies indicating that the structures are dynamically stable. We found that the structures maintained the tetrahedral bonding environment similar to AlN, however the large lattice mismatch ($\sim$17\%) between AlN and \emph{w}-BN led to large distortion and rotation of these tetrahedra. The unit cells of the \baln themselves had space group symmetries that differed from that of the wurtzite lattice, and they were largely the same for each structure at boron fraction $x$ and $1-x$. The structural deviations of the \baln from the wurtzite structure were quantified in terms of the average bond lengths, bond angles, and the 'tetrahedrality' of the structure. Average bond lengths and bond angles were calculated using the cartesian coordinates of each atom and their nearest neighbors; and averages were computed over all possible bond lengths and angles in the conventional unit cell. This 'tetrahedrality' score was computed for all the alloys based on the bond angles, bond lengths, and solid angle distributions at all the sites for each \baln structure \cite{Pan2021,Zimmermann2020}. A tetrahedrality score of 1 indicates a perfect tetrahedral bonding environment and 0 indicates a completely non-tetrahedral bonding environment. Our calculations showed that the tetrahedral distortion maximizes near $x=0.6$, along with the formation energies, and is especially high in the $x=0.583$ and $x=0.6$ ground-state structures \cite{Milne2023}. Detailed analysis and description of the structural properties including tetrahedrality are available in our previously published article \cite{Milne2023} and images of each structure are available in the accompanying Supplementary Information.

\subsection{Electronic Properties}
\subsubsection{Bandgaps}
\label{band gaps}
Table \ref{table:bandgaps} compares our $GW_0$ and DFT computed bandgaps of \emph{w}-AlN, \emph{w}-BN, and the \baln alloys with those reported in the literature. Since \emph{w}-BN exists only as a high-pressure metastable phase with no existing experimental measurements of its bandgap, we compare to existing theoretical prediction its bandgap and also make comparisons for the other polymorphs of BN for which experimental measurements exist, i.e.,  hexagonal (\emph{h}-BN) and cubic (\emph{c}-BN). As expected, the DFT computed bandgaps are much smaller than those predicted from the $GW_0$ simulations. We find that $GW_0$ calculated values are in excellent agreement with experimental values found in the literature. For \emph{w}-BN the values are in good agreement with previously reported computed bandgaps. We find that the partial self-consistent $GW_0$ calculation (6.19 eV) is necessary to reproduce the experimental bandgap of AlN (6.2 eV) \cite{Shen2017,Kudrawiec2020}, as the one-shot $G_0W_0$ calculation underestimates the bandgap for AlN to be $5.51$ eV. A comparison of $G_0W_0$ and $GW_0$ computed bandgap for all the \baln alloys can be found in Figure S1 and Table S1 in the Supplementary Information. 

\begin{table*}[]
    \centering
    \begin{tabular}{ c c c c c } 
    \hline
  & DFT-PBE & $GW_0$ & Calc. & Exp. \\ \hline
\emph{w}-AlN & 4.06  & 6.19 & 6.18 \cite{Shen2017}$^b$, 6.2 \cite{Zhang2017}$^a$ & 6.1 \cite{Lawniczak-Jablonska2000}, 6.12--6.19 \cite{Shen2017},                       \\
 & & & & 6.25 \cite{Kudrawiec2020}
\\
\emph{w}-BN & 5.39 & 7.47 & 6.39 \cite{Ahmed2007}$^c$ 6.8 \cite{Zhang2017}$^a$, &  --   \\
 & & & 6.84 \cite{Shen2017}$^b$, 5.44--7.70 \cite{Kudrawiec2020} , 8.06$^d$\cite{Gao2012}&
\\
\emph{c}-BN & 4.45 & 6.52 & 5.43 \cite{Ahmed2007}$^c$, 7.13$^d$\cite{Gao2012} & 6.0, 6.1 \cite{Ahmed2007}, 6.1--6.4 \cite{Lawniczak-Jablonska2000}        \\
\emph{h}-BN & 4.23 & 6.34 & 4.83 \cite{Ahmed2007}$^c$, 5.98 \cite{Shen2017}$^b$ & 5.4--5.8 \cite{Ahmed2007, Lawniczak-Jablonska2000}, 6.08 \cite{Cassabois2016}  \\  
\emph{w}-B$_{0.25}$Al$_{0.75}$N & 4.47 & 6.52 & ~5.4 \cite{Zhang2017}$^a$ & -- \\
\emph{w}-B$_{0.5}$Al$_{0.5}$N & 4.95 & 6.83 & ~5.8 \cite{Zhang2017}$^a$ & -- \\
\emph{w}-B$_{0.75}$Al$_{0.25}$N & 5.15 & 7.14 & ~6.5 \cite{Zhang2017}$^a$ & -- \\
\hline
\end{tabular}
    \caption{$GW_0$ and DFT computed bandgaps of AlN, three polymorphs of BN, and three \baln structures compared with existing experimental and computed values reported in the literature. \\ \footnotesize{$^a$ LDA with scissor operation, $^b$ HSE, $^c$ GGA form proposed by Engel and Vosko (GGA-EV), $^d$ QPscGW}}
    \label{table:bandgaps}
\end{table*}

\begin{center}
\begin{figure}[h]
\centering
  \includegraphics[]{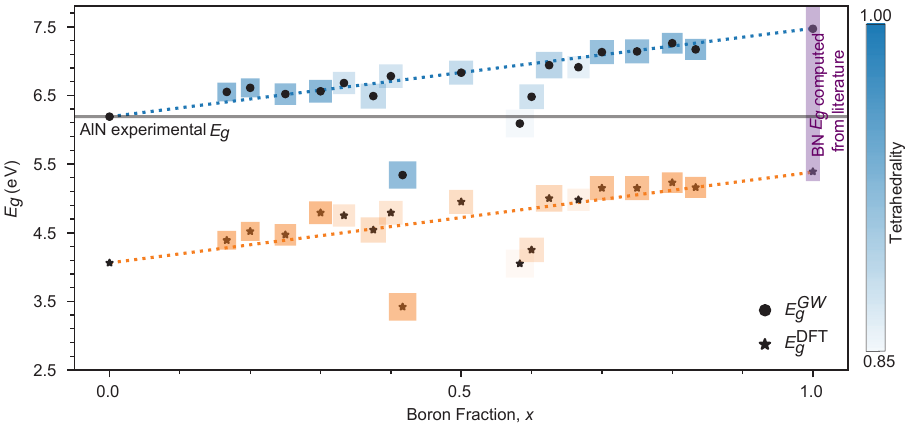}
  \caption{Black circle and star symbols indicate $GW_0$ and DFT computed bandgaps, respectively, of the \baln structures, $w$-AlN, and $w-$BN. The sizes of the blue and orange squares are proportional to the formation energy of the structures  \cite{Milne2023}. Their transparency is proportional to the average tetrahedrality of the \baln \cite{Milne2023} with the tetrahedrality scale denoted by the color bar. The solid gray horizontal line denotes the experimental bandgap of AlN (6.2 eV) \cite{Kudrawiec2020} and the purple bar denotes the range for computed bandgaps of \emph{w}-BN \cite{Kudrawiec2020}. The dotted lines show Vegard's law fits for the $GW_0$ and DFT bandgaps excluding the bandgaps at $x=0.417$, 0.583, and 0.600.}
  \label{fig:bandgap_vs_x}
\end{figure}
\end{center}

Figure \ref{fig:bandgap_vs_x} shows $GW_0$ and DFT computed bandgaps of all the \baln structures. The $GW_0$ and DFT gaps are shown by circle and star symbols, respectively.  The size of the overlaying boxes is proportional to the formation energies of the structures and their color transparency denotes the tetrahedrality of the structure. With the exception of three outlying boron fractions, at $x=0.417$, 0.583, and 0.6, the bandgap of \baln alloys increases almost linearly with increasing $x$. The dotted lines in Figure \ref{fig:bandgap_vs_x} show Vegard's law fits for the $GW_0$ and DFT bandgaps. When the outlying bandgaps at $x$=0.417, 0.583, and 0.600 are excluded from the fit, the root-mean-square error (RMSE) for the $GW_0$ bandgaps and the DFT bandgaps are 0.088 eV and 0.15 eV, respectively. Vegard's law fits that include all the data points yield a very large RMSE value of 0.40 eV and 0.38 eV for $GW_0$ and DFT, respectively. 

At $x=0.417$ the anomalously low bandgap can be attributed to the presence of $sp^2$ bonding in one of the B-atoms in the lattice. For the outliers at $x=0.583$ and $x=0.6$, the reason for low bandgaps is not immediately obvious, but we find that these structures are among the ones with the lowest tetrahedrality score, and thus a high distortion in tetrahedral bonds and high standard deviation of N-Al-N bond angles. Thus, it is expected that experimentally grown high-quality \emph{w}-\baln alloys will display an increase in bandgap as the boron content increases. However, the presence of polycrystalline phases with mixed $sp^3$ and $sp^2$ bonded BN and/or defects could dramatically alter the bandgap with respect to the pure wurtzite structure. 

\subsubsection{Element and orbital projected band structures}
The bandstructures of \baln alloys can not be directly compared with that of \emph{w}-AlN or \emph{w}-BN bandstructures since the \baln structures are supercells of the primitive wurtzite lattice, albeit with small deviations from the wurtzite lattice. In the \baln bandstructure, the electronic bands of the primitive wurtzite-like lattice get folded into the smaller Brillouin zone of the supercell. Thus we unfolded the \baln bandstructures onto the wurtzite primitive cell Brillouin zone and obtained the spectral weights which give the probability that a supercell band has the same Bloch character as the primitive cell band \cite{Medeiros2014}. Further, we examined the element and orbital contributions to the spectral weights by projecting their elemental contributions as in Refs. \cite{Ikeda2017, Schuler2018}. This analysis was restricted to the DFT bandstructures for computational tractability. Note that the nature of the bands remains practically invariant between DFT and $GW_0$ because the wavefunctions from DFT change negligibly at the $GW_0$ level of correction \cite{Biswas2023}. The folded $GW_0$ bandstructures of all the \baln can be found in Figures S2--S5 in the Supplementary Information.

\begin{center}
\begin{figure}
    \centering
  \includegraphics[width=0.30\linewidth]{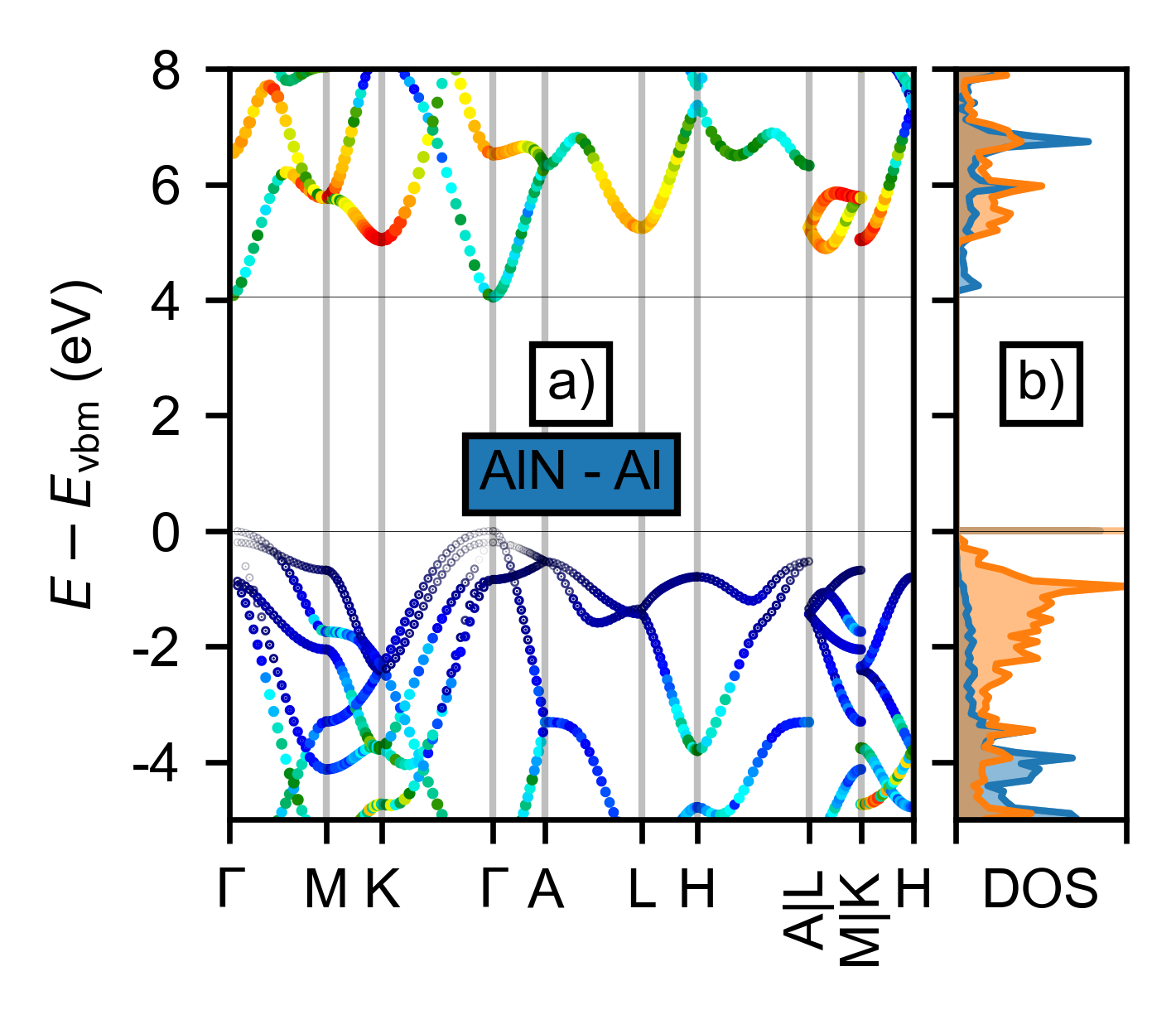}
 \includegraphics[width=0.30\linewidth]{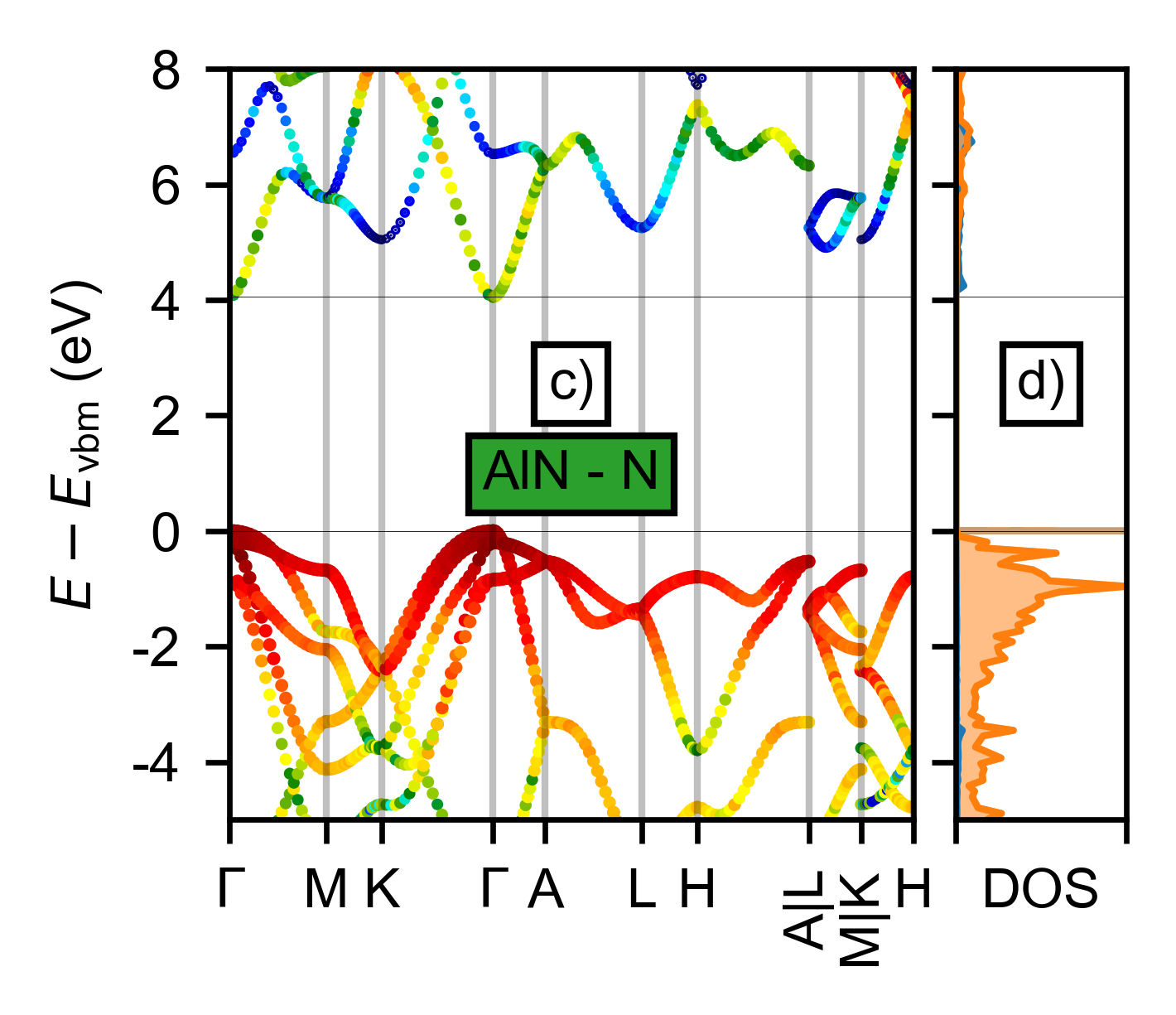}\\
  \includegraphics[width=0.30\linewidth]{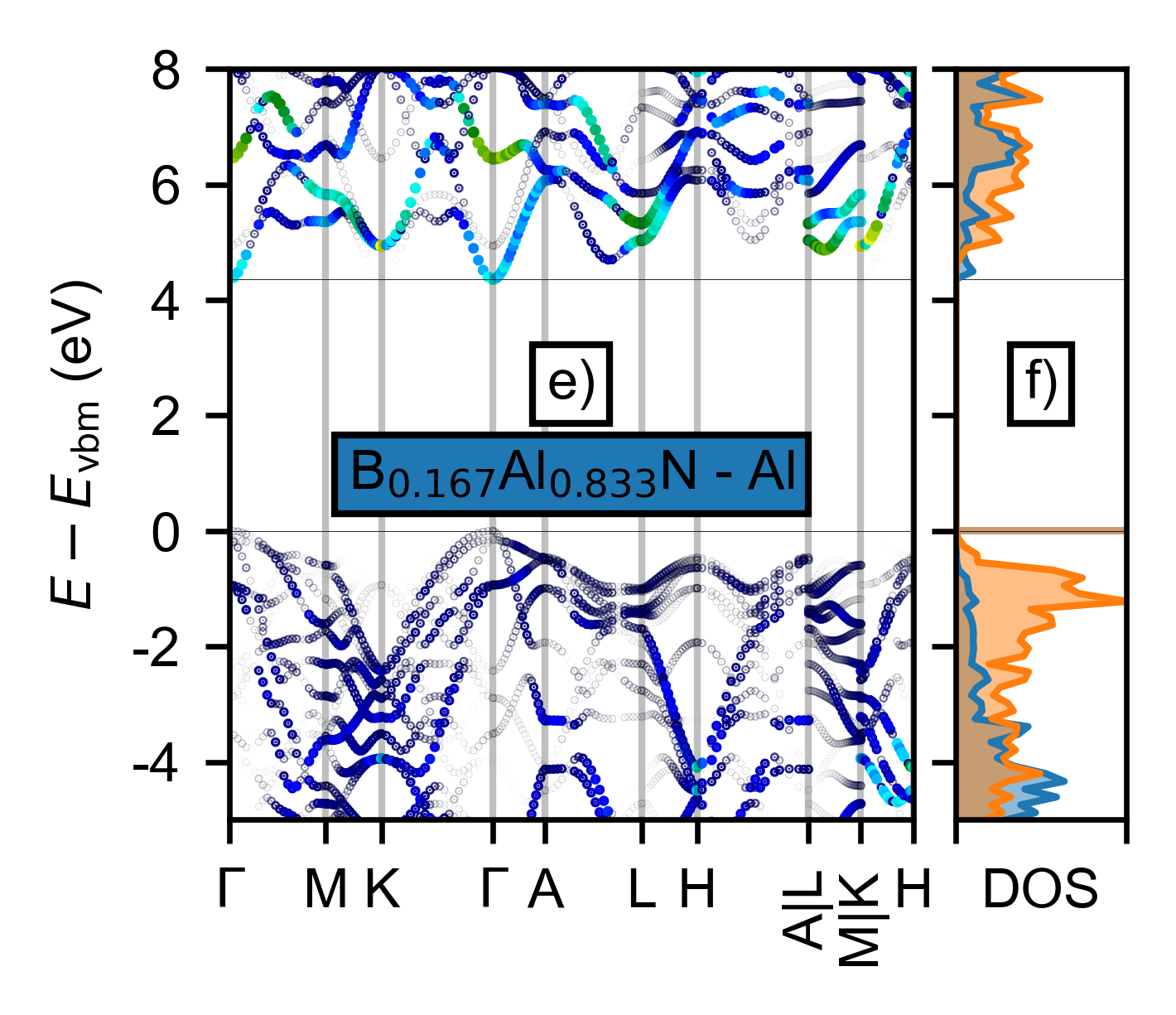}
  \includegraphics[width=0.30\linewidth]{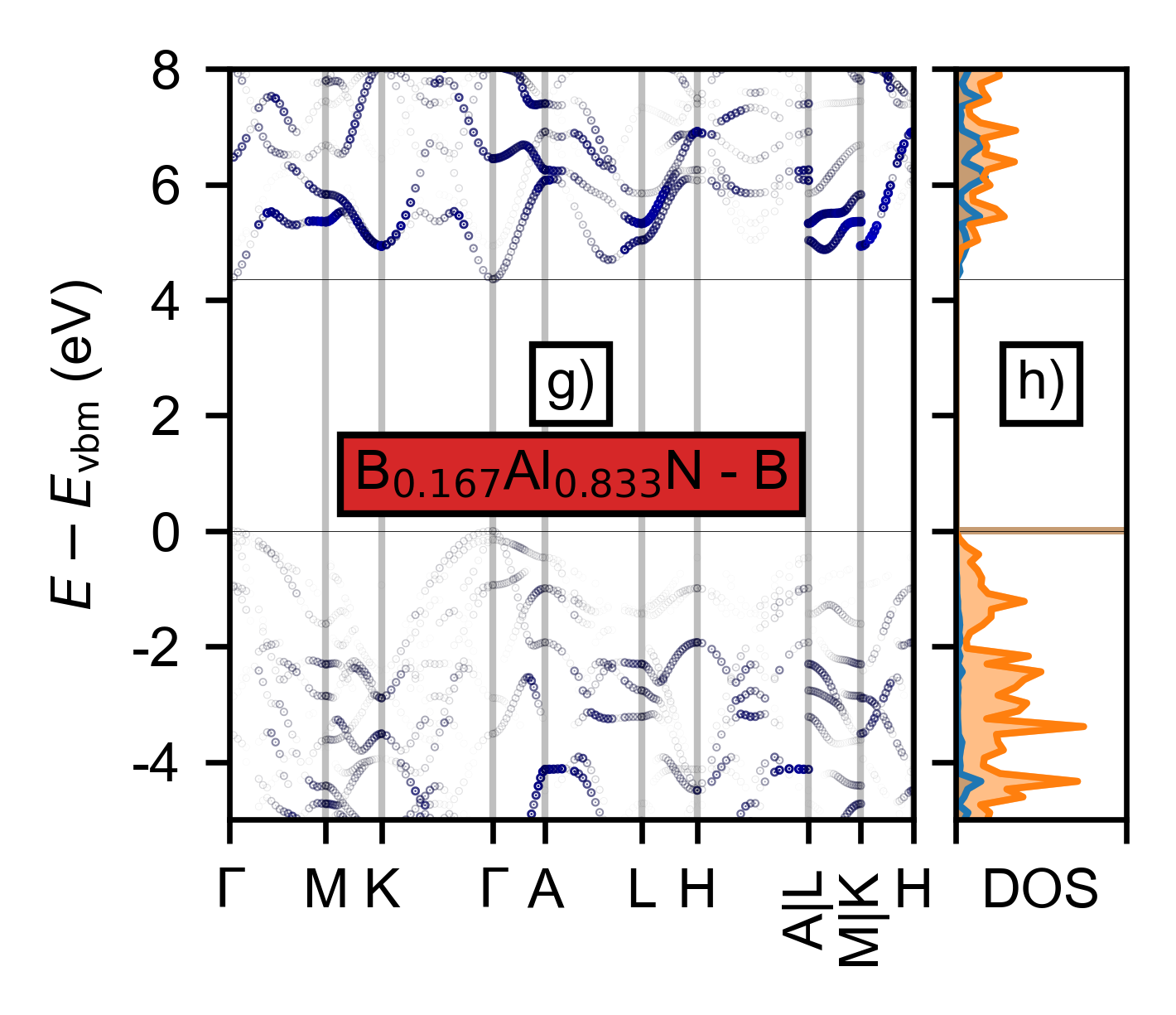}
  \includegraphics[width=0.30\linewidth]{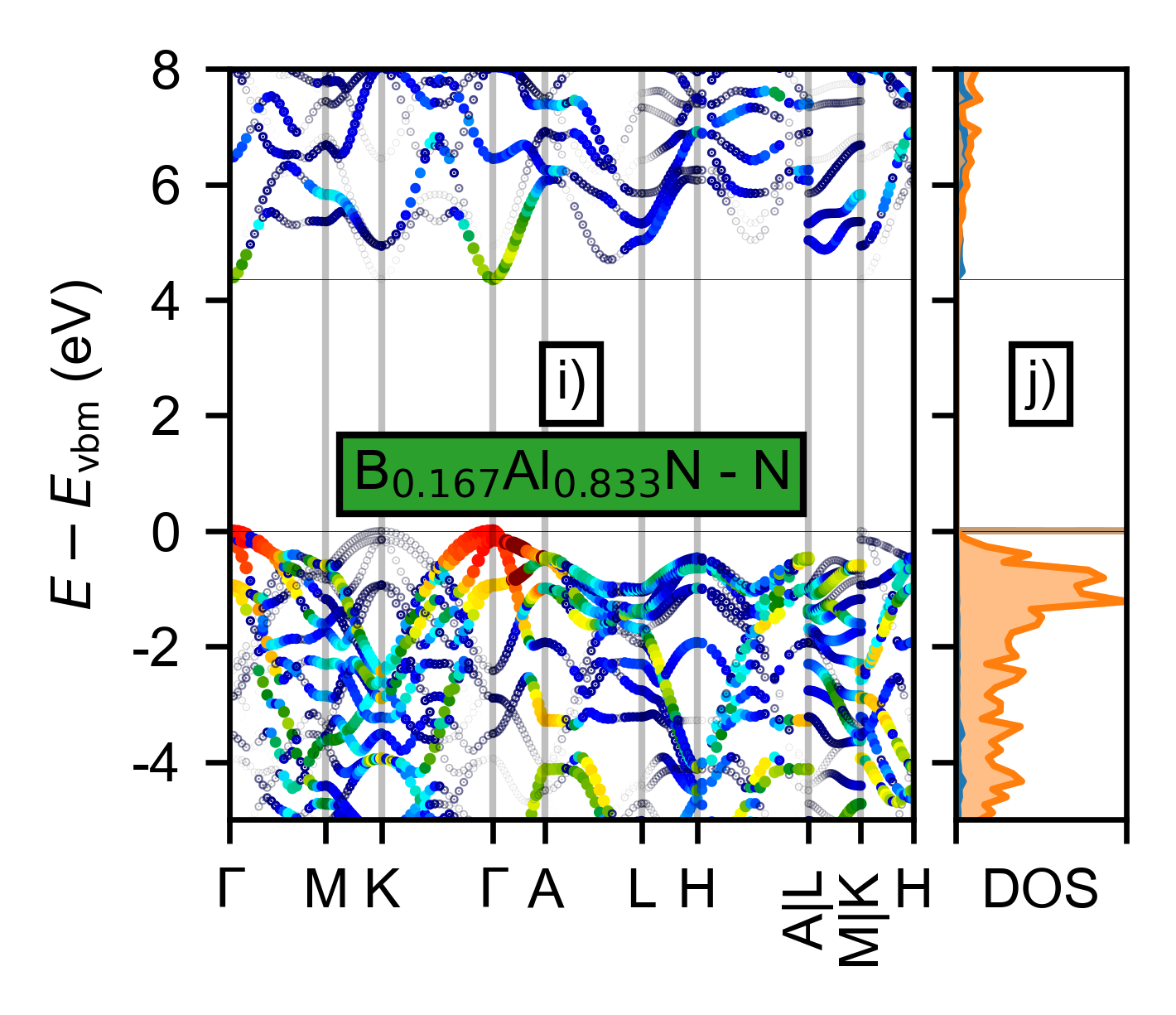}\\
 \includegraphics[width=0.30\linewidth]{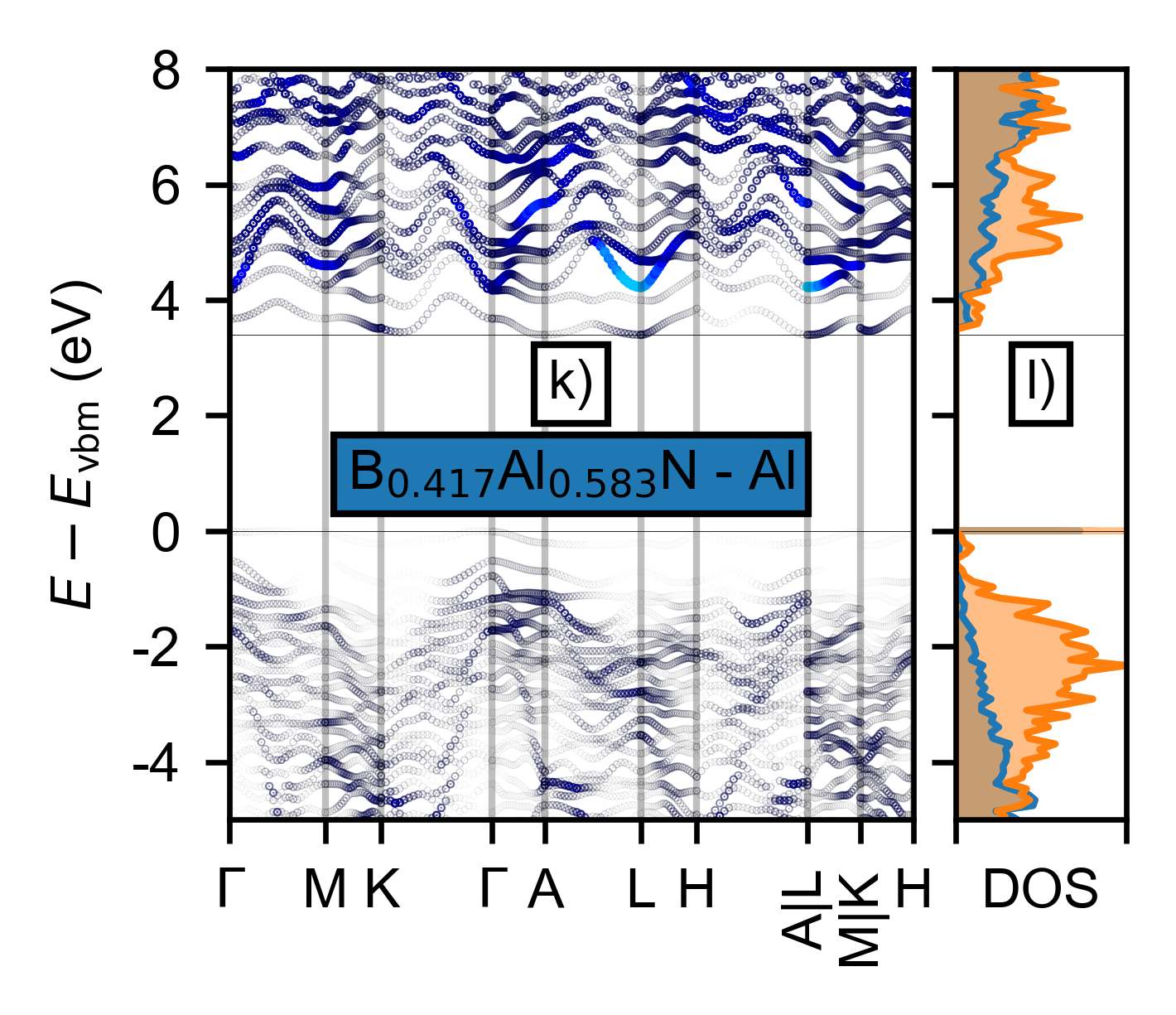}
  \includegraphics[width=0.30\linewidth]{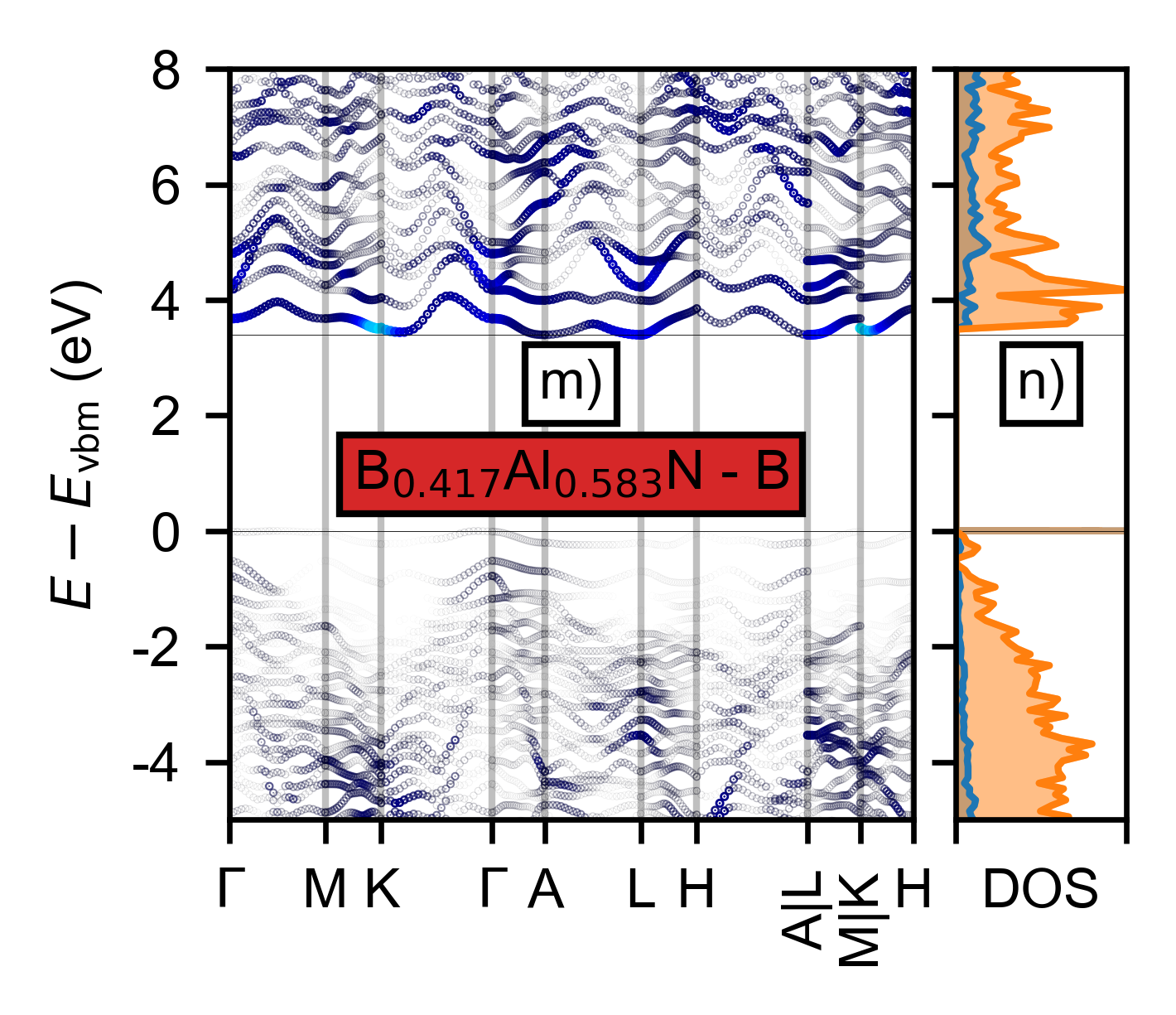}
  \includegraphics[width=0.30\linewidth]{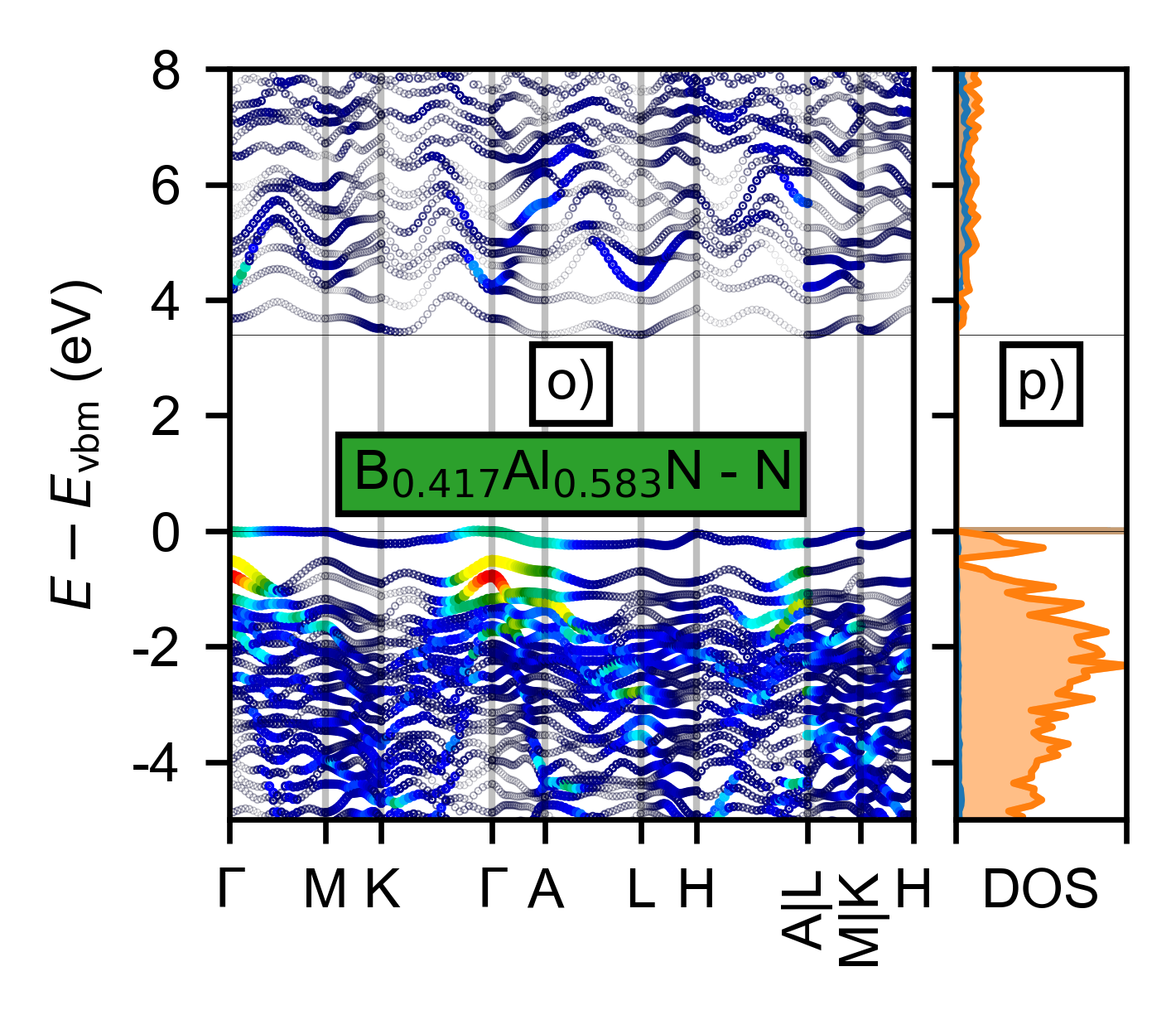}\\
  \includegraphics[width=0.30\linewidth]{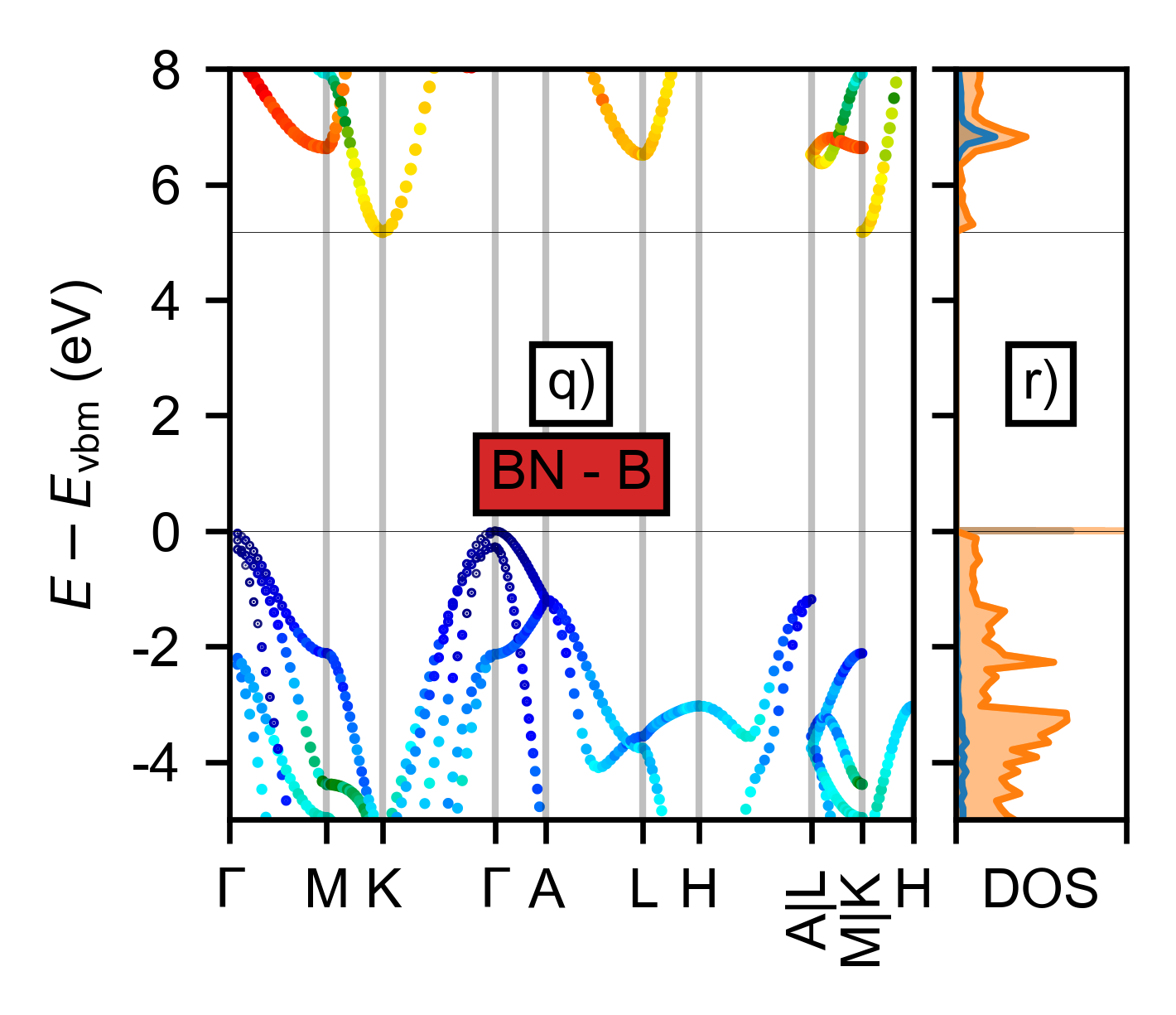}
  \includegraphics[width=0.30\linewidth]{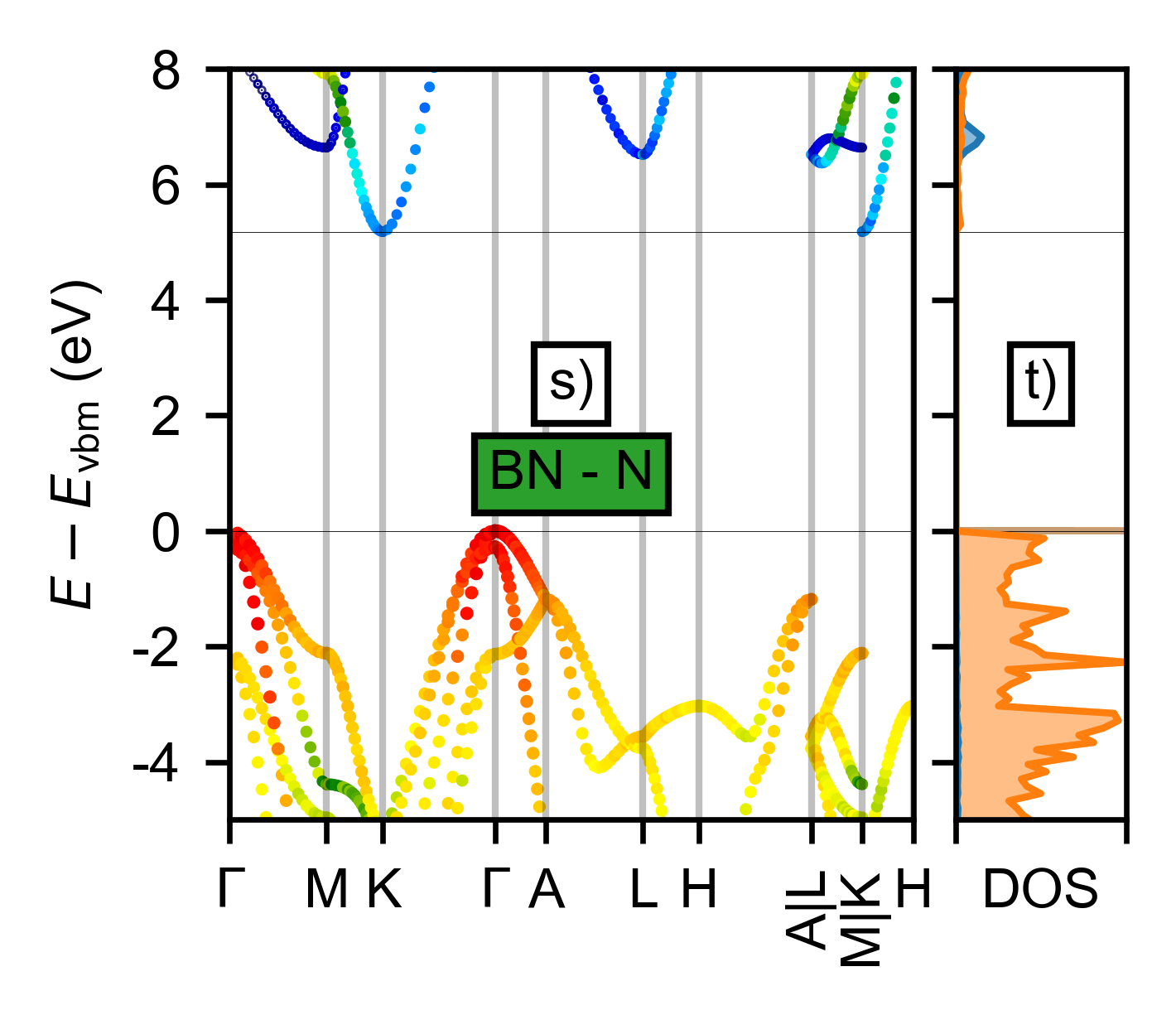}\\
  \includegraphics[]{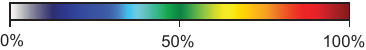}
    \caption{Element-projected and unfolded DFT bandstructures and the orbital-projected DOS for AlN, B$_{0.167}$Al$_{0.833}$N, B$_{0.417}$Al$_{0.583}$N, and \emph{w}-BN. The colorbar represents the elemental contribution to the number of bands in the primitive cell, i.e. the spectral weight, and ranges from 0\% to 100\%. In the DOS, the blue lines represent the contribution from $s$ states and the orange lines represent the contribution from all the $p$ states. The DOS are in arbitrary units. $E_{\mathrm{vbm}}$ is the energy of the valence band maximum, which is subtracted from the band structure such that the VBM is at 0 eV.}
  \label{fig:pbanddpos}
\end{figure}
\end{center}

Figure \ref{fig:pbanddpos} shows the unfolded and element-projected DFT bandstructures and the orbital-projected densities of states (DOS) of \emph{w}-AlN, B$_{0.167}$Al$_{0.833}$N, B$_{0.417}$Al$_{0.583}$N, and \emph{w}-BN.  The unfolded element projected bandstructures and the orbital-resolved densities of states of all other \baln alloys are available in the Supplementary Materials Figures S6--S10.  

Figures \ref{fig:pbanddpos}a-d show that AlN exhibits a direct bandgap at the $\Gamma$-point. Figure \ref{fig:pbanddpos}a shows that the Al-atom contributions to the bands are dominant in the L-- and K--valleys. At the conduction band in the $\Gamma$--valley, the lowest energy bands are found to be dominated by the $s$ orbitals of N atoms, see Figure \ref{fig:pbanddpos}d. Figures \ref{fig:pbanddpos}q-t show that \emph{w}-BN exhibits a substantially higher bandgap, 7.47 eV, in comparison to AlN and is also indirect in nature from $\Gamma$--K. While \emph{w}-BN has a similar energy and contribution from the cation $p$ states in the K-- and L--valley as AlN, the conduction band $\Gamma$--valley has much higher energy, which causes the primary difference in the \emph{w}-BN and AlN bandstructures, namely a direct bandgap in AlN and an indirect bandgap in \emph{w}-BN. 

In Figures \ref{fig:pbanddpos}e-j, we can see that the low B-fraction alloy, B$_{0.167}$Al$_{0.833}$N, maintains a direct bandgap at $\Gamma$ and the N-dominated conduction band $\Gamma$--valley behavior is similar to that observed for AlN. Interestingly, the valence band minima (VBM) is not substantially altered due to the incorporation of B, but both the K-- and the L--valleys are lower in energy for the conduction bands. The B-contribution is found to be predominant near K, M, and L--valleys in the lower conduction bands. 

Figure \ref{fig:pbanddpos}k-p shows the band structure for the $x=0.417$ structure, which we noted earlier for its drastically diminished bandgap and presence of $sp^2$ bonded B-sites. We find two flatter low-lying bands near the conduction band minima. They are primarily due to the boron atoms. A substantial contribution is from the $sp^2$ orbitals of B-atoms that are observed in two of the five B-atoms in this structure \cite{Milne2023}. The contribution of the $sp^2$ and $sp^3$ bonded B-sites and N-sites to the bandstructures are available in the Supplemental Materials Figure S11. Additionally, two flatter valence bands can be seen, which are a result of $sp^2$ bonded nitrogen. Above the two low-lying conduction bands, one can see a dominant L--valley, and the $\Gamma$--valley. Consequently, the low bandgap can be directly linked to the two $sp^2$ bonded boron sites in this structure. This finding is particularly noteworthy as it suggests that the existence of $sp^2$ bonded boron in \emph{w}-\baln has the potential to significantly reduce the bandgap of \emph{w}-\baln and increase the effective masses of holes and electrons.

\subsubsection{Direct-to-indirect transition}
 \begin{center}
\begin{figure}[h]
   \centering
  \includegraphics[width=0.5\linewidth]{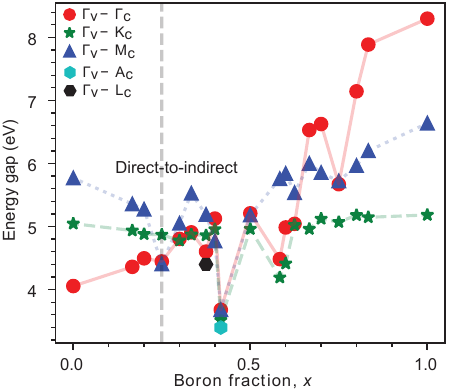}
  \caption{The energy gaps at $k_\mathrm{V}-k'_\mathrm{C}$ where $k_\mathrm{V}$ and $k'_\mathrm{C}$ are the location of the VBM and CBM $k$-points in the Brillouin zone, respectively. The $\Gamma_\mathrm{V}$--$\Gamma_\mathrm{C}$ are denoted by red circles, $\Gamma_\mathrm{V}$--K$_\mathrm{C}$ by green stars, and $\Gamma_\mathrm{V}$--M$_\mathrm{C}$ by blue triangles. $\Gamma_\mathrm{V}$--A$_\mathrm{C}$ are shown by cyan hexagons and $\Gamma_\mathrm{V}$--L$_\mathrm{C}$ by black hexagons for structures where the gaps at $\Gamma_\mathrm{V}$--A$_\mathrm{C}$ and $\Gamma_\mathrm{V}$--L$_\mathrm{C}$ are lower than those at $\Gamma_\mathrm{V}$--M$_\mathrm{C}$ or $\Gamma_\mathrm{V}$--K$_\mathrm{C}$. The dashed gray line shows the direct-to-indirect gap transition which is located between $x=0.20$ and $x=0.25$.}
  \label{fig:bandgap_direct_vs_x}
\end{figure}
\end{center}
Figure \ref{fig:bandgap_direct_vs_x} shows the energy gap of all the \baln alloys for transitions between the $\Gamma$, M, and K \emph{k}-points. Energy gaps are also shown for the A and L \emph{k}-points when the energy gaps at these \emph{k}-points are lower than all the other energy gap values. The energy gap transitions are denoted as $k_\mathrm{V}-k'_\mathrm{C}$ where $k_\mathrm{V}$ and $k'_\mathrm{C}$ are the location of the VBM and CBM $k$-points in the Brillouin zone, respectively. All the unfolded bandstructures that were used to obtain the energy gaps at the various $k_\mathrm{V}$ and $k'_\mathrm{C}$ are available in the Supplemental Materials Figures S6--S10. 

Figure \ref{fig:bandgap_direct_vs_x} shows that for all \baln the VBM is located at the $\Gamma$-point. The location of the CBM is at the $\Gamma$-point for $x<0.250$, for $0.250 \le x < 0.417$ the location varies, and for $x > 0.417$, the CBM is clearly at the K-point. Thus we can distinctly see that a direct-to-indirect transition occurs at $x=0.25$. Previous studies that considered \baln structures formed by random cation substitution in the AlN-lattice report the direct-to-indirect gap transition in \baln alloys to be between $x=0.12$ and $x=0.28$ \cite{Shen2017, Zhang2017, Akiyama2018}, but one study reported it to be at $x=0.66$ \cite{Dong2021}. As found in our work, Shen et al. \cite{Shen2017} and Zhang et al. \cite{ZhangQ2022} also reported CBM to be at the K-point for higher fractions of B.

Interestingly, the energy gap at $\Gamma_\mathrm{V}-\Gamma_\mathrm{C}$ rises steeply with $x$, whereas the energy gap at $\Gamma_\mathrm{V}-\mathrm{M}_\mathrm{C}$ and $\Gamma_\mathrm{V}-\mathrm{K}_\mathrm{C}$ remain relatively constant. One can see the outlying bandgaps at $x=0.417$, $0.583$, and $0.600$, which can be attributed to the high tetrahedral distortion in the bonding environments of their respective structures, as discussed in Section \ref{band gaps}.

\subsubsection{Effective Masses}

Table \ref{table:effective_masses} shows the effective masses of \emph{w}-AlN and \emph{w}-BN. The heavy hole effective masses of AlN are 2.99 $m_e$ and they decrease to 1.11 $m_e$ for \emph{w}-BN. These values agree well with a previous report by Zhang et al. \cite{Zhang2017}. The effective masses of AlN and \emph{w}-BN show only a slight anisotropy in terms of their electron, heavy hole, and light hole effective masses. This has also been observed in the other binary III-nitrides (AlN, GaN, and InN) \cite{Punya2012}. On the other hand, the electron effective masses stay relatively constant in the range 0.40 -- 0.60 $m_e$. Therefore, one may expect that \baln alloy effective masses will be in this range. Tables S2 and S3 in the SI show that this is largely the case, however, there are outliers that can be attributed to the high tetrahedral distortion in the intermediate alloy structures, which can distort the bandstructure from the expected wurtzite symmetry. Thus, a clear trend with increasing B-fraction is not observed in our lattices since all the alloys have different deviations from the ideal wurtzite lattice. Furthermore, the local strain on the B-N bonds and Al-N bonds vary between the alloy structures (see our previous article \cite{Milne2023}) which are likely to have an impact on the valence band character. Additionally, the very high effective masses seen in the $x=0.417$ structure can be explained by the presence of $sp^2$ bonding that causes bands with very low dispersion near the Fermi level. These effective masses do not take into account the spin-orbit interaction, which is known to have a small effect on the predicted effective mass bands in wurtzite AlN\cite{Kim1997}. A comparison of the bandstructures near the band edges with and without the effects of spin-orbit coupling is available in the Supplementary Information Figures S14 and S15.

\begin{table}[h]
    \centering
    \begin{tabular}{ c c c } 
    \hline
Hole effective masses ($-m_e$) & AlN & \emph{w}-BN \\
\hline
$\Gamma$--M             & 2.96 & 1.45 \\
$\Gamma$--K             & 2.99 & 1.11 \\
$\Gamma$--A             & 0.27 & 1.22 \\
\hline
Electron effective masses ($m_e$) & & \\
$\Gamma$--M             & 0.55 & - \\
$\Gamma$--K             & 0.45 & - \\
$\Gamma$--A             & 0.40 & - \\
K--M                    & - & 0.52 \\
K--A                    & - & 0.61 \\
\hline
\end{tabular}
    \caption{Effective masses for AlN and \emph{w}-BN structures are shown which are taken from parabolic fits to the band edges.}
    \label{table:effective_masses}
\end{table}

\subsection{Static dielectric constants}
\begin{center}
\begin{figure}[h]
\centering
 \includegraphics[]{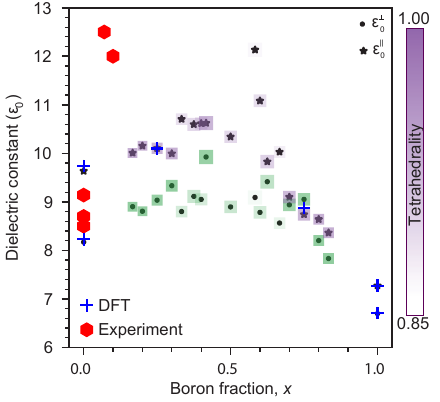}
  \caption{Static dielectric constants, $\varepsilon^{\perp}_0$ (black circles) and $\varepsilon^{\parallel}_0$ (black stars) of \baln, \emph{w}-AlN, and \emph{w}-BN. Red hexagons show experimentally measured dielectric constants of AlN \cite{Collins1967, Liufu1998, Yu2021}, B$_{0.07}$Al$_{0.93}$N and B$_{0.10}$Al$_{0.90}$N \cite{Zhu2021,Hayden2021} reported in the literature. DFT computed values obtained from the Materials Project database \cite{MaterialsProject} are shown as blue plus symbols. The size of green and lavender squares is proportional to the formation energies of the \baln and indicates the $\varepsilon^{\perp}_0$ and $\varepsilon^{\parallel}_0$, respectively. The transparency of the squares is proportional to the tetrahedrality of \balnend, where the color bar indicates the tetrahedrality value.}
  \label{fig:dielectric}
\end{figure}
\end{center}

Figure \ref{fig:dielectric} shows the total static dielectric tensor components of the \balnend, \emph{w}-AlN, \emph{w}-BN. The out-of-plane dielectric constants, $\varepsilon_{0}^{\perp}$, are shown as black circles and are defined as, $\varepsilon_{0}^{\perp}=\frac{\varepsilon^{xx}_0 + \varepsilon^{yy}_0}{2}$. The in-plane dielectric constants, $\varepsilon_{0}^{\parallel}$, are indicated by black star symbols and are equivalent to $\varepsilon^{zz}_0$. 

We can see that some \baln structures have much larger dielectric constants than the constituent bulk materials, \emph{w}-AlN=9.6$~\varepsilon_{0}$ or \emph{w}-BN=7.3$~\varepsilon_{0}$, with values reaching as high as 12.1$~\varepsilon_{0}$. The static dielectric constant is known to scale as the summation over $i$ of $\Tilde{Z^2}/\lambda_i^2$  \cite{zhao2002first}, where $\Tilde{Z}$ is the mode effective charge, and $\lambda$ is $i_{\mathrm{th}}$ infra-red (IR) active phonon mode. Thus, the dramatic increases in $\varepsilon_{0}^{\perp}$ and $\varepsilon_{0}^{\parallel}$ values could be due to two possible effects, 1) an increase in ionicity due to a change in local charge distribution, or 2) a decrease in structural stability and softening of IR-active phonon modes. Figure S12 in the Supplementary Information shows a charge isosurface comparison between \emph{w}-AlN and B$_{0.167}$Al$_{_0.833}$N, indicating a clear increase in ionicity in the B-N bonds in the [0001] direction. A more detailed analysis of the vibrational, dielectric, and piezoelectric properties of these alloys is necessary to determine how these effects result in the dramatic increase in lattice dielectric constant in \balnend. 

We also observe large bowing in $\varepsilon_{0}^{\perp}$ and $\varepsilon_{0}^{\parallel}$ with bowing parameters of $b \approx -7.2~\varepsilon_0$ and $b \approx -9.8~\varepsilon_0$, respectively. In contrast, in the case of other group-III-nitride like \algan and In$_x$Al$_{1-x}$N \cite{Kafi2020,Ambacher2021} the bowing is insignificant. However, some alloys do exhibit bowing trends similar to that of \balnend, for example the highly mismatched Sc$_x$Al$_{1-x}$N \cite{Ambacher2021}, wurtzite Zn(O, S), and wurtzite Zn(O, Se) alloys \cite{Baldissera2016}. 

We also compute the high-frequency dielectric constants or the electronic part of the dielectric constants. Interestingly, the high-frequency dielectric constants remain nearly invariant across the boron fraction and are between 4 to 5~$\varepsilon_0$. All high-frequency dielectric constants can be found in Supplementary Materials Figure S13.  

There are very few existing reports of the dielectric constants of \balnend. DFT predicted dielectric constants of B$_{0.25}$Al$_{0.75}$N (mp-1019380) and B$_{0.75}$Al$_{0.25}$N (mp-1019379) are available in the open-source Materials Project (MP) database \cite{MaterialsProject}. While the MP values, shown in Figure \ref{fig:dielectric} as blue pluses, are in good agreement with those obtained in our study, the lattices and atomic arrangements in the MP structures differ from those studied in this work. 

A recent study by Zhu et al. \cite{Hayden2021, Zhu2021} measured the permittivity of epitaxially grown low boron content ($x=0.07, 0.10$) \baln and showed that they have high relative permittivity at 50 K (12 $\varepsilon_0$ for $x=0.10$ and 12.5 $\varepsilon_0$ for $x=0.07$) with low dielectric loss as a function of temperature at 10 kHz, 100 kHz, and 1 MHz, substantially higher than AlN ($9.6 \varepsilon_{0}$). Similar permittivity, 12.2 $\varepsilon_0$, has been reported in polycrystalline B$_{0.10}$Al$_{0.90}$N films \cite{Liljeholm2011}. The dielectric constants computed in our study are not as high as those seen experimentally \cite{Hayden2021,Zhu2021,Liljeholm2011}, but we do observe a substantial increase in dielectric constant at intermediate boron fractions of \balnend. We postulate that this disagreement between theory and experimentally measured values could be due to the high strain in the fabricated thin films due to the lattice-mismatch with substrate, defects, grain boundaries, and/or mixed phases. 

\subsection{Electric breakdown fields}
Experimental data of the breakdown field, $E_{b}$, in \baln alloys are scarce in the literature, however, a breakdown field of $\sim$ 6.4 MV/cm with weak temperature dependence has been measured in B$_{0.07}$Al$_{0.93}$N  \cite{Zhu2021}. Traditionally, first-principles simulations with empirical deformation potentials have been applied to theoretically estimate the intrinsic breakdown field in semiconductors and insulators. Recently, fully \emph{ab initio} methods of calculating the dielectric breakdown field have been developed, which involve computationally expensive DFPT calculations of the electron-phonon scattering rates \cite{Sun2012}. Here, we employ a machine-learned model developed by Kim et al.  \cite{Kim2016} to predict the breakdown field at a vastly reduced computational cost. This model, Equation \ref{eq:electric_breakdown_fit}, requires bandgap energy and $\Gamma$-point phonon frequencies to estimate the $E_{b}$. We utilize the bandgaps obtained from our $GW_0$ simulations and the $\Gamma$-point phonon frequencies from the phonon spectra reported in our previous work \cite{Milne2023}. 
\begin{center}
\begin{figure}[h]
\centering
  \includegraphics[]{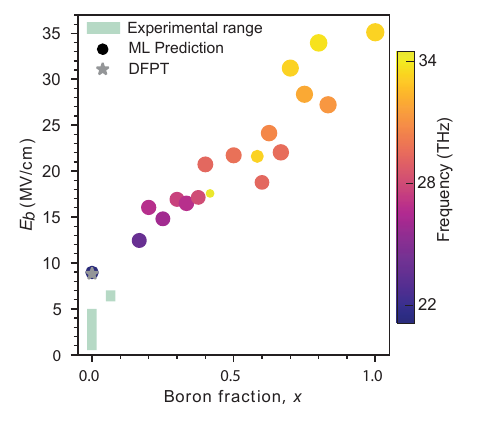}
  \caption{Electric breakdown fields, $E_{b}$, of \baln, $\emph{w}$-AlN, and $\emph{w}$-BN are shown as circles and were computed using a model by Kim et al. \cite{Kim2016}. The size of the circles is proportional to the $GW_0$ bandgap energy. Their color denotes the magnitude of the $\Gamma$-point phonon frequency and is noted by the color bar. The green bar represents the experimentally measured range of $E_{b}$, including measurements done at AC and DC voltages, for AlN \cite{Ruemenapp1999, An2005, Abdallah2011, Liufu1998, Gaskins2017} and B$_{0.07}$Al$_{0.93}$N \cite{Zhu2021}. The star symbols indicates the fully \emph{ab initio} prediction of the $E_{b}$ for AlN \cite{Kim2016}.}
  \label{fig:breakdown}
\end{figure}
\end{center}

Figure \ref{fig:breakdown} shows the $E_{b}$ of \baln, \emph{w}-AlN, and \emph{w}-BN as a function of $x$. The $E_{b}$ for \baln alloys ranges from 9.0 MV/cm to 35 MV/cm. The predictions of the AlN $E_{b}$ are in excellent agreement with fully \emph{ab initio} prediction of the $E_{b}$ for AlN from Kim et al. \cite{Kim2016}. Note that the experimentally measured values of $E_{b}$ of AlN \cite{Ruemenapp1999, Liufu1998, An2005, Abdallah2011, Gaskins2017}, shown as a green band in Figure \ref{fig:breakdown} are substantially lower, 1--5 MV/cm, than those predicted in this work, 8.8 MV/cm. Experimentally measured $E_{b}$ vary heavily in the literature for a given material, due to the varying growth process and the different types of experimental measurements \cite{Sun2012, ColtrinAlGaN2017, Ruemenapp1999}. Thus, experimentally measured $E_{b}$ in \baln may be much lower than those shown in Figure \ref{fig:breakdown} due to the presence of defects and impurities. Therefore, our results provide a theoretical maximum for what could be achieved in wurtzite \balnend. It is worth noting that the predicted $E_{b}$ of \baln are extremely high $E_{b}$'s, much higher than has been seen in diamond (10--21.5 MV/cm) \cite{Sun2012,Slobodyan2022}, which has one of the highest measured breakdown fields. 

We can also see a nearly linear dependence of $E_{b}$ with $x$. Interestingly, the reduced bandgap in some of the \baln structures also correlates with an increase in the predicted phonon cutoff frequency. This causes the \baln which exhibited outlying bandgaps to follow a linear trend for the $E_{b}$. Applying a linear regression fit to our electric breakdown values, the linear dependence can be described by:
\begin{equation}
    E_{b}(x) = 25.71 x + 8.47~\mathrm{MV/cm},
\label{eq:electric_breakdown_fit}
\end{equation}
where $x$ is the boron fraction in \balnend. 

\section{Conclusion}
In summary, we predict the electronic properties, dielectric properties, and breakdown fields of the recently predicted 17 ground states of \baln in
the $x = 0-1$ range using first-principles density functional theory and many-body perturbation theory within $GW$ approximation. The structures used in this study are the lowest formation energy configurations predicted by a DFT-based cluster expansion \cite{Milne2023}, which samples a large phase space, including non-wurtzite lattices. Thus these structures are more likely to be observed experimentally than the previously studied structures that are predominantly obtained by substitution of B in the wurtzite-AlN lattice \cite{Dong2021,Zhang2017,Shen2017, ZhangQ2022, Ota2022, Lu2021}. Moreover, it is noteworthy that we have consiered a significantly higher number of boron fractions (17) in comparison to the studies reported in the literature.
In this article, we have made bandgap and bandstructure predictions of \baln alloys using the \emph{GW} method. This is the first study that employs the excited state many-body perturbation theory \emph{GW} method for the \baln system. This method is known to provide better agreement with experimental bandgaps for a variety of large gap materials \cite{Punya2012, Bakhtatou2016, Gao2012, vanSchilfgaarde2006,Chen2014}. We find that the bandgaps of \baln vary linearly from that of $\emph{w}$-AlN (6.19 eV) to that of $\emph{w}$-BN (7.47 eV). We observe few outliers to the trend where \baln with any $sp^2$-bonded B or low tetrahedrality display reduced bandgap. Unfolded element-projected bandstructures show that a direct-to-indirect gap transition occurs at $x \approx 0.25$, which agrees well with previous predictions in the literature. We find that hole effective masses decrease overall as a function of B-fraction, while the electron effective masses do not vary substantially. We see that the presence of $sp^2$ bonding can lead to localized states with high effective masses. Furthermore, our work is the first study to report DFT simulated dielectric constants of \balnend, revealing that they exhibit large dielectric constants that bow heavily as a function of the boron content. Additionally, we make the first predictions for electric breakdown fields in \balnend--electric breakdown fields predicted using the model of Kim et al. \cite{Kim2016}, show a linear increase in electric breakdown field, from 9.0 to 35 MV/cm as the boron content increases. Thus, \baln alloys present an exciting opportunity for tunable gap, dielectric, and high-breakdown field materials for a variety of applications. 
\\
\\
\\

\textbf{Supplementary Materials} \par 
Supplementary Materials are available from Materials Today or from the author. Our Supplementary Materials provide more information on the $G_0W_0$ bandgaps, $GW_0$ and elemental projected bandstructures, effective masses, and dielectric constants.

\textbf{Acknowledgements} \par 
This work was supported by ULTRA, an Energy Frontier Research Center funded by the U.S. Department of Energy (DOE), Office of Science, Basic Energy Sciences (BES), under Award No. DESC0021230. The authors acknowledge the San Diego Supercomputer Center under the NSF-XSEDE and NSF-ACCESS Award No. DMR150006, the NSF-FuSE Award No. 2235447, and the Research Computing at Arizona State University for providing HPC resources. This research used resources of the National Energy Research Scientific Computing Center, a DOE Office of Science User Facility supported by the Office of Science of the U.S. Department of Energy under Contract No. DE-AC02-05CH11231. The authors have no competing interests to declare. 

\bibliography{main}

\end{document}